\documentclass[a4paper]{jpconf}
\usepackage{graphicx}
\usepackage{lineno}
\begin{document}
\title{Anisotropy studies with the Pierre Auger Observatory}

\author{Jo\~ao de Mello Neto$^1$, for the Pierre Auger Collaboration$^2$}

\address{$^1$Instituto de F\'isica, Universidade Federal do Rio de Janeiro, Ilha do Fund\~ao\\
Rio de Janeiro, RJ  Brazil \\
$^2$Observatorio Pierre Auger, Av. San Mart\'in Norte 304, 5613 Malarg\"ue, Argentina \\
(Full author list: http://www.auger.org/archive/authors\_2012\_04.html)}

\ead{jtmn@if.ufrj.br}


\begin{abstract}
We report recent results from the Pierre Auger Observatory about the study of  the anisotropy in the arrival directions of  ultra-high energy cosmic rays. We present the results of the search for a dipolar anisotropy at the EeV energy scale.  Measurements of the  phase and the amplitude of the first harmonic modulation in the right-ascension distribution are discussed.  For cosmic rays with energies above 55 EeV, we present an update of the search for correlations between their arrival directions and the positions of active galactic nuclei from the V\'eron-Cetty and V\'eron catalog.   
We also discuss the results of correlation analyses applied to other populations of extragalactic objects.
Finally we present the search for  anisotropies in the data without the usage of astronomical catalogues. 
\end{abstract}

\section{Introduction}
After almost 50 years since the first detection \cite{Linsley:1963km} of cosmic rays of the order of 100 EeV, their origin and nature remain unknown. For energies up to $10^{15}$ eV, cosmic rays are believed to have a galactic origin and shock acceleration in supernova remnants could be the most likely source. At the highest energies, the most probable sources of  ultra-high energy cosmic rays (UHECRs)  are extragalactic:   jets of active galactic nuclei (AGN), radio lobes, gamma ray bursts and colliding galaxies, among others \cite{kotera}.
Measurements of anisotropies in the arrival directions of UHECRs, together with a detailed study of their energy spectrum and their  mass composition, are essential to their understanding.  One clear cluster of arrival directions around a given celestial object would point to a source and measurements of the spectrum and other observables related to the source would be possible. Even if clear single sources are not isolated in the data, a class of sources could be deduced from a pattern of arrival directions, for instance through the correlation of the arrival directions of UHECR with populations of astrophysical objects listed in catalogues.  Another  measurement especially relevant for galactic cosmic rays would be a detection of a large scale anisotropy, such as a dipole, a quadrupole or other pattern.

The Pierre Auger Observatory \cite{observ} is the world's largest UHECR detector, conceived to measure the flux, arrival direction distribution and mass composition of cosmic rays from 0.1 EeV (low energy extensions included \cite{heat,amiga}) to the very highest energies with high statistical significance. It is a hybrid detector that combines both surface and fluorescence detection at the same site.  The surface detector (SD) consists of 1660 10m$^2$ $\times$ 1.2\,m   water-Cherenkov stations  deployed over 3000
km$^2$ on a 1500 m triangular grid plus a denser, `infill" array (on a 750~m grid). The SD is overlooked by a fluorescence detector (FD) composed of five fluorescence buildings with a total of 27 telescopes. The surface detector   stations
sample at the ground level the charged particles in the shower front. The fluorescence telescopes  can record the ultraviolet 
light emitted as the shower
crosses the atmosphere. 
The fluorescence detector operates only on clear, moonless nights, so its duty cycle is about 13\%. On the other hand, the  surface detector array has a duty cycle close to 100\%. 

\section{Search for large scale anisotropy}

The physical origin of the structure in the cosmic ray spectrum known as  the ``ankle"  (at  about 4 EeV)  is  not settled yet. There are two main candidate scenarios to explain it.
In the first one, the ankle is related to the transition from a galactic component to a harder extragalactic one \cite{linsleyproc,hillas}. In the second one (so called dip scenario),  cosmic rays are supposed to be extragalactic protons down to energies below 1 EeV and the concave shape is due to the effect of energy losses of protons by pair creation with cosmic microwave background photons (CMB) \cite{berez}.   In order to elucidate this issue, anisotropy measurements are very important.
If the ankle is due to the change from a galactic to an extragalactic component, a dipolar modulation in the  cosmic ray arrival directions is expected. The amplitude of the dipole should increase with energy up to the ankle, reaching a level of a few percent, even if  primaries are heavy nuclei. Predictions of the amplitude and orientation of the dipole depend on the magnitude of the magnetic field and also on the source distribution. If the dip scenario is the correct explanation for the ankle, it means that UHECRs above $10^{18}$ eV are already dominated by the extragalactic component, and their flux is expected to be highly isotropic. But a dipole of less than one percent amplitude is expected due to the {\it Compton-Getting effect}. It would produce a dipolar distribution constant with energy and with an orientation related to the relative movement of the Earth with respect to the "rest frame"  of the extragalactic cosmic rays. 
\begin{figure}[h]
\begin{minipage}{18pc}
\includegraphics[width=18pc]{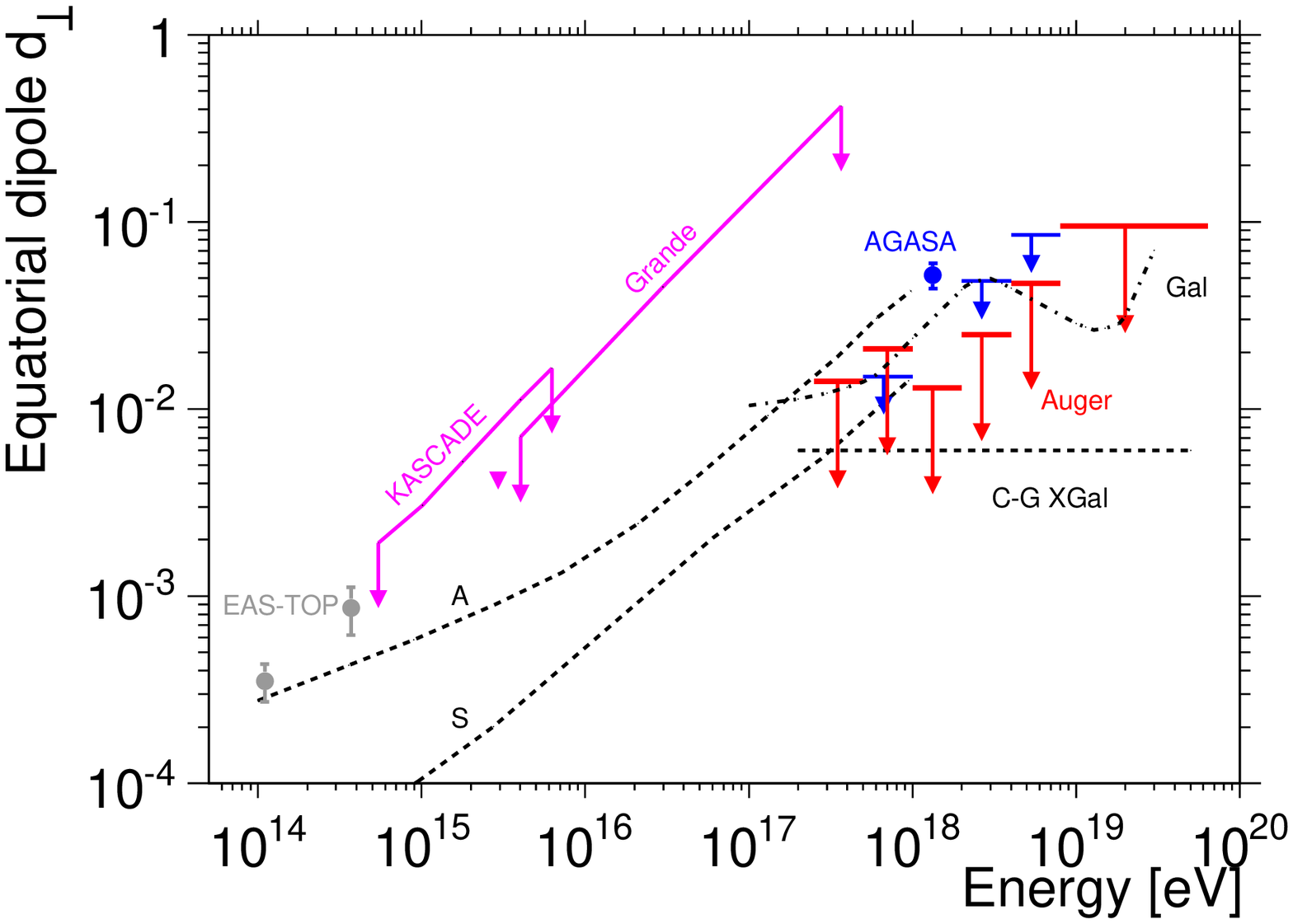}
\caption{\label{fg1a}   Upper limits on the anisotropy amplitude of first harmonic as a function of energy and model predictions (see text). }
\end{minipage}\hspace{2pc}%
\begin{minipage}{18pc}
\includegraphics[width=18pc]{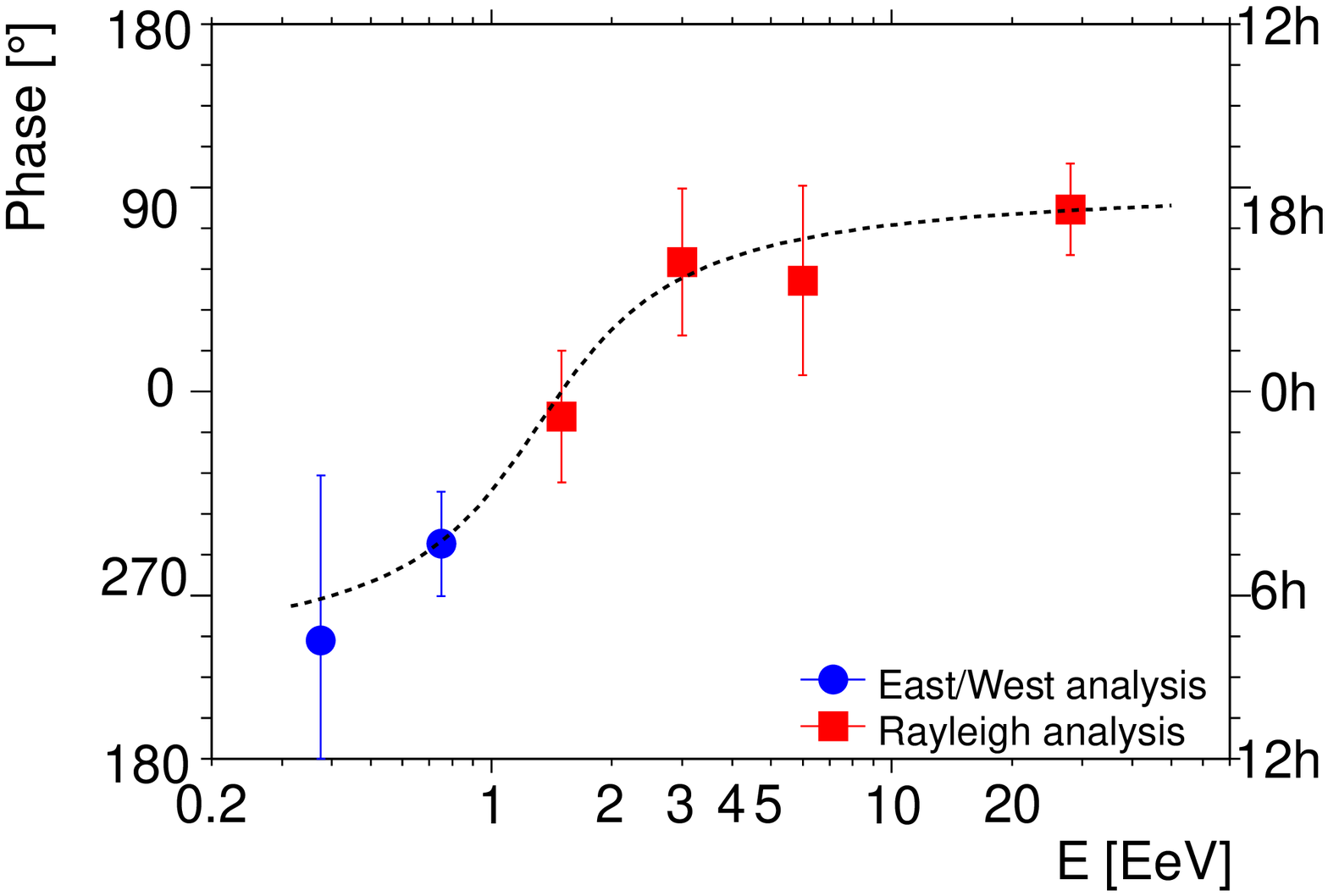}
\caption{\label{fg1b} Phase of the first harmonic as a function of energy. The dotted line represents a hypothetical smooth transition of dipole direction. }
\end{minipage} 
\end{figure}
Thanks to the high statistics of the SD data a first harmonic analysis was performed in different energy ranges starting from $2.5\times 10^{17}$ eV in search for dipolar modulations in right ascension \cite{dipole,lyberis}.  This analysis used the fact that the exposure is almost uniform in right ascension.
No significant amplitude was detected. The 99\% CL upper limits as a function of energy are shown in  Fig.~\ref{fg1a} together with predictions   from two different galactic magnetic field models (A and S),  for a purely galactic origin of UHECRs   (Gal), and the expectations from the Compton-Getting effect (C-G Xgal) \cite{dipole}.  The particular model with an antisymmetric halo magnetic field (A) is already excluded by the upper limits. In Fig.~\ref{fg1b} the phase measurement as a function of the energy shows an interesting pattern: it suggests a smooth transition between a phase of $\sim 270^{\circ}$ (consistent with the right ascension of the galactic center) below $1\times 10^{18}$ eV and another phase  of  $\sim 90^{\circ}$ (consistent with the right ascension of the galactic anti-center) above $5\times 10^{18}$~eV. This is interesting since a real anisotropy would need less events to be established with high statistical confidence from the phase consistency in ordered energy intervals than  by amplitude measurements  \cite{dipole}. New data will show if this feature still stands.

\section{Correlation with celestial objects}

\begin{figure}[h]

\begin{minipage}{17pc}
\includegraphics[width=17pc]{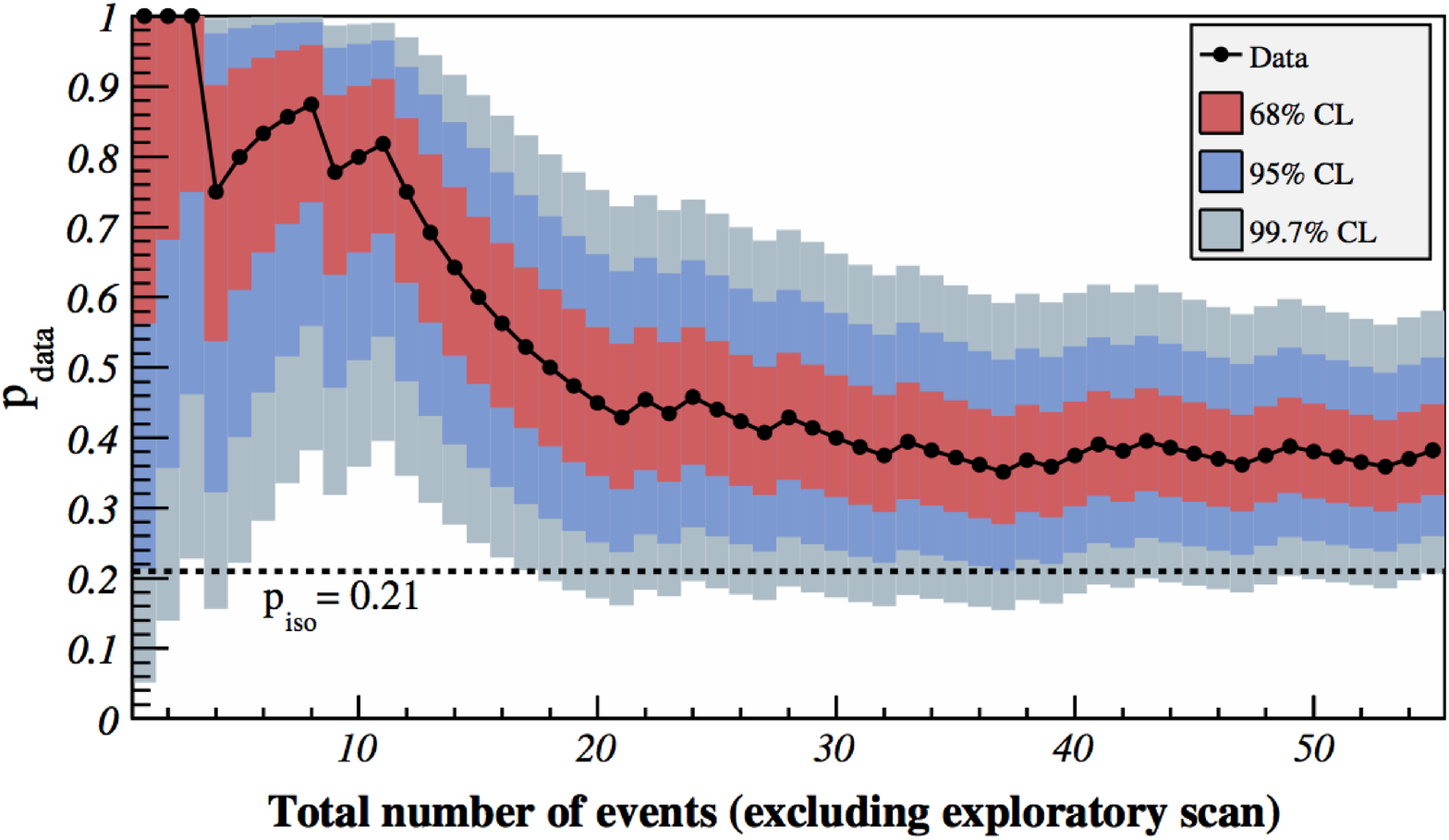}
\caption{\label{fg3} Fraction of events correlating with AGNs as a function of the cumulative number of events, starting after the exploratory data. The expected correlating fraction for isotropic cosmic rays is shown by the dotted line. }
\end{minipage} \hspace{2pc}%
\begin{minipage}{17pc}
\includegraphics[width=17pc]{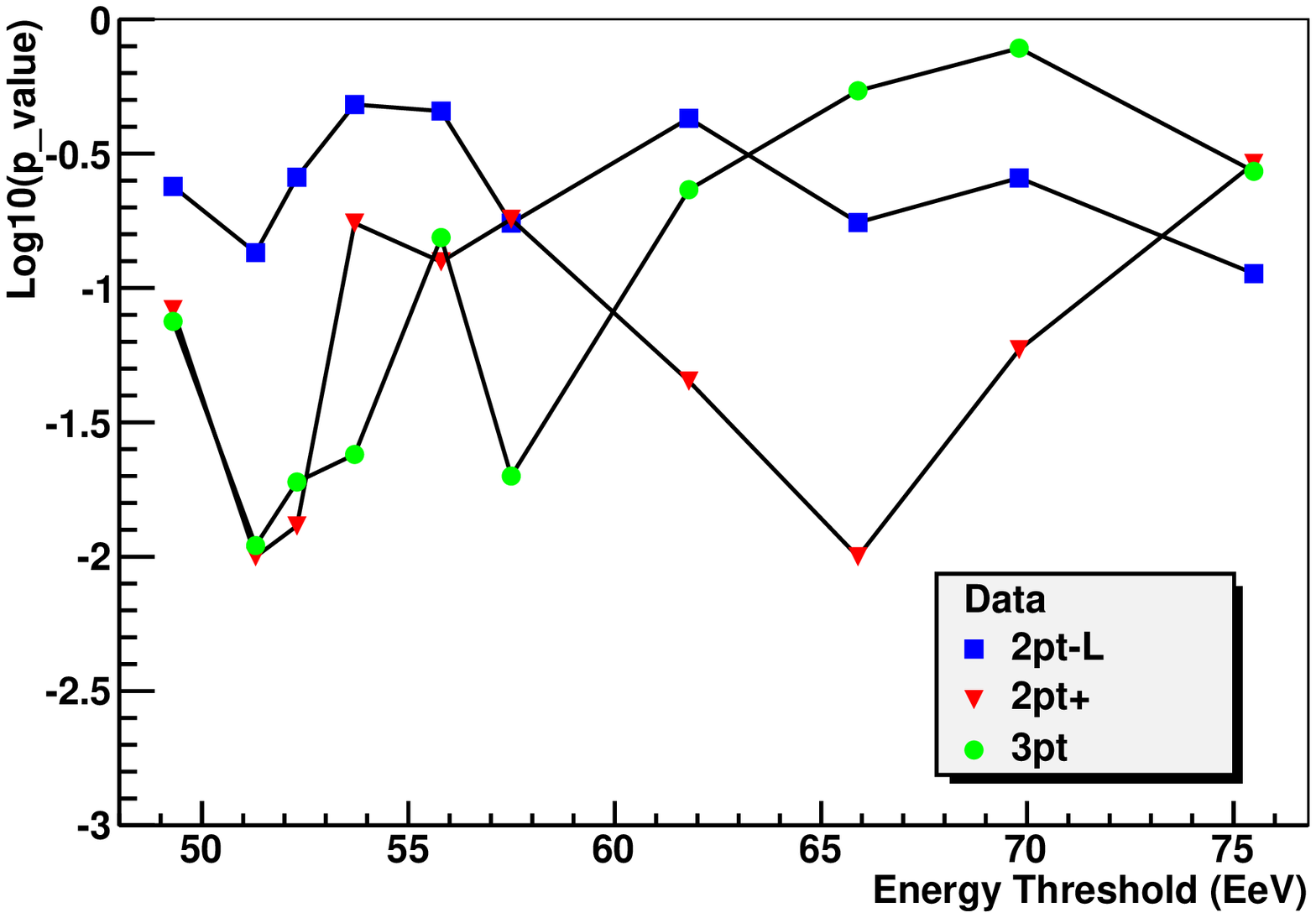}
\caption{\label{fg4}  $P_{\mathrm{value}}$ of the Auger data for the 2pt-L, 2pt+ and 3pt methods. 
The minimum in $P_{\mathrm{value}}$ is at 100 events for the 2pt+ and 3pt methods and corresponds to 
an energy of about $51\approx$ EeV .  \hfil \break  } 
\end{minipage}
\end{figure}

For the highest energy cosmic rays, possible candidates able to accelerate them up to $10^{20}$~eV are jets of AGN, as mentioned before. 
The Pierre Auger Collaboration reported \cite{agn, science} a correlation of events with the AGNs in the VCV catalogue. The first 14 events were used for an exploratory scan  that yielded the following search parameters: energy threshold ($E_{th} = 55$ EeV), maximum angular separation ($\Psi \leq 3.1^{\circ}$) and maximum redshift ($z\leq 0.018)$. Those parameters minimize the probability that the correlation with AGN could result from a background fluctuation if the flux were isotropic. The subsequent 13 events established a 99\% confidence level for rejecting the hypothesis of isotropic cosmic ray flux. The reported fraction of correlation events was ($69^{+11}_{-13})$\%. 
By adding data with $E_{th} = 55$ EeV up to the end of 2009 (69 events in total), the amount of correlation decreased to $(38^{+7}_{-6})$\% \cite{update}. For this dataset, we show in  Fig.~\ref{fg3}  the most likely value of the fraction of the correlated events plotted with black dots as a function of the total number of time-ordered events (the events used in the exploratory scan are excluded).   The most updated estimate of the fraction of correlating cosmic rays is $(33\pm 5)$\%, with 21\% expected under the isotropic hypothesis \cite{kampertICRC}.  
                                                                                                                                                                                                                                                                                                                                                                                                                                                                                                                                                                                                                                                                                                                            
{\it A posteriori }  studies \cite{update}  showed that the distribution of arrival directions of  the 69 cosmic rays is compatible with models (for suitable values of two parameters, the smoothing factor $\sigma$ and an isotropic fraction $f_{\textrm{iso}}$) based on  populations of nearby extragalactic objects, such as galaxies in the 2MRS and AGNs in the SWIFT-BAT catalogues. The models fit the data for smoothing angles around a few degrees and for correlating fractions of order 40\% ($f_{\textrm{iso}} \approx 0.6$). 
The data does not fit either the isotropic expectation or the predictions of the models with  $f_{\textrm{iso}}= 0$. 
A large isotropic fraction could indicate that the model is not using a catalogue that includes all the contributing UHECR sources or that a fraction of arrival directions are isotropized by large magnetic deflections due to large electric charges.

\section{Intrinsic anisotropy}
Anisotropy can also be searched with the use of catalog independent methods. In Ref.~\cite{intrinsic} three methods named 2pt-L, 3pt and 2pt+, each giving a different measure of self-clustering in arrival directions, were tested on mock  data sets. The impact of sample size and magnetic smearing was studied. These studies suggested that the three methods could efficiently respond to a real anisotropy in the data with a $P_{\mathrm{value}}$ = 1\% or smaller with a data set of the order of one hundred events.
The methods were applied to the data from the 20 highest energy events to the highest 110. As shown in Fig.~\ref{fg4}, the two most powerful methods (3pt and 2pt+) show a minimum in the distribution of $P_{\mathrm{value}}$ as a function of the energy threshold for an energy around 51 EeV,  (i.e., at 100 events) \cite{intrinsic}.
 But there is no $P_{\mathrm{value}}$ smaller than 1\% in any of the 30 (correlated) scanned values. There is thus no strong evidence of clustering in the data set which was examined. 
In case there is a true weak  anisotropy in the data, the common minimum   could  indicate  the onset of this anisotropy,  while the less powerful method (2pt-L) would have not detected it. For higher energy thresholds the number of events decreases and the power of the methods diminishes as expected.    Simulations show \cite{intrinsic} that this null result is compatible with the correlation with the VCV catalogue discussed before.

\section{Conclusion}                                                                                                                    
 The Pierre Auger Observatory has detected high quality events and has made key measurements of the highest-energy cosmic rays. In spite of that, many issues remain open.  The measurements of the large scale anisotropy provided upper limits in the dipole amplitude that constrain theoretical models in the ankle region.  
The Auger data provide evidence for a correlation between arrival directions of cosmic rays above 55 EeV and the positions of AGNs with $z < 0.018$. The fraction of correlating cosmic rays has decreased to about 33\%  compared with 21\% expected for isotropy. 
Searches for anisotropy using catalogue independent methods were performed and no statistically significant signal was found.

 \textbf{}
\section{References}



\begin{thebibliography}{00}  



\bibitem{Linsley:1963km} Linsley J 1963  {\it Phys. Rev. Lett.} {\bf 10} 146
 
\bibitem{kotera} Kotera K and Olinto A V 2011 {\it Annu. Rev. Astron. Astrophys.} {\bf 49} 119

\bibitem{observ}  Abraham J {\it et al.}  (The Pierre Auger Collaboration) 2001 {\it Nucl. Instr. and Meth. C\/} {\bf A 523}  50

\bibitem{heat}  Hermann-Josef Mathes T   (The Pierre Auger Collaboration) 2011 {\it Proc. 32nd ICRC}  ({\it Preprint} 1107.4807)

\bibitem{amiga} Wundheiler B (The Pierre Auger Collaboration) 2011 {\it Proc. 32nd ICRC}  ({\it Preprint} 1107.4807)

\bibitem{linsleyproc}  Linsley J  1963 {\it Proc. 8th ICRC}   {\bf 4} 77 

\bibitem{hillas} Hillas A M  1967 {\it Phys. Lett A} {\bf 24} 677

\bibitem{berez} Berezinsky V {\it et al.} 2006  {Astron. Astrophys.} {\bf 043005} 74 
  
\bibitem{dipole}  Abraham J {\it et al.}  (The Pierre Auger Collaboration) 2011  {\it Astropart. Phys.} {\bf 34}  627
 
\bibitem{lyberis} Lyberis H (The Pierre Auger Collaboration) 2011 {\it Proc. 32nd ICRC}  ({\it Preprint} 1107.4805)

\bibitem{agn} Abraham J {\it et al.}  (The Pierre Auger Collaboration) 2008 {\it Astropart. Phys.} \textbf{29} 188  

\bibitem{science}  Abraham J {\it et al.}  (The Pierre Auger Collaboration) 2007 {\it Science} \textbf{318} 939

\bibitem{update} Abraham J {\it et al.}  (The Pierre Auger Collaboration) 2010 {\it Astropart. Phys.} \textbf{34} 314
 
\bibitem{kampertICRC} Kampert  K-H  (The Pierre Auger Collaboration) 2011 {\it Proc. 32nd ICRC}  ({\it Preprint} 1207.4823)

\bibitem{intrinsic}   Abreu P {\it et al.}  (The Pierre Auger Collaboration) 2012 {\it J. Cosmol. Astropart. Phys.} JCAP04(2012)040

\end{thebibliography}
\end{document}